\documentclass[aps,prl,twocolumn,groupedaddress]{revtex4-2}

\usepackage{bm}
\usepackage{float}
\usepackage{booktabs,tabularx,array}
\usepackage{graphicx,color}
\usepackage[colorlinks=true,allcolors=blue]{hyperref}%

\usepackage{enumerate}
\usepackage{harpoon}
\usepackage{amsmath,amssymb}
\usepackage{gensymb}
\usepackage{accents}

\definecolor{mmcolor}{rgb}{0.8, 0.0, 0.2}

\definecolor{ctcolor}{rgb}{0.8, 0.0, 0.2}

\newcommand{\vecn}[1]{\vec{#1}}
\newcommand{\vecd}[1]{\bm{#1}}

\begin{document}
\title{Generic elasticity of thermal, under-constrained systems}
\author{Cheng-Tai Lee}
\affiliation{Aix Marseille Univ, Universit\'e de Toulon, CNRS, CPT (UMR 7332), Turing Center for Living Systems, Marseille, France}
\author{Matthias Merkel}
\email[]{matthias.merkel@cnrs.fr}
\affiliation{Aix Marseille Univ, Universit\'e de Toulon, CNRS, CPT (UMR 7332), Turing Center for Living Systems, Marseille, France}


\begin{abstract}
	Athermal (i.e.\ zero-temperature) under-constrained systems are typically floppy, but they can be rigidified by the application of external strain, which is theoretically well understood.
	Here and in the companion paper, we extend this theory to \emph{finite} temperatures for a very broad class of under-constrained systems.
	In the vicinity of the athermal transition point, we derive from first principles expressions for elastic properties such as isotropic tension $t$ and shear modulus $G$ on temperature $T$, isotropic strain $\varepsilon$, and shear strain $\gamma$, which we confirm numerically.
	These expressions contain only three parameters, entropic rigidity $\kappa_S$, energetic rigidity $\kappa_E$, and a parameter $b_\varepsilon$ describing the interaction between isotropic and shear strain, which can be determined from the microstructure of the system.
	Our results imply that in under-constrained systems, entropic and energetic rigidity interact like two springs in series.
	This also allows for a simple explanation of the previously numerically observed scaling relation $t\sim G\sim T^{1/2}$ at $\varepsilon=\gamma=0$.  
	Our work unifies the physics of systems as diverse as polymer fibers \& networks, membranes, and vertex models for biological tissues.
\end{abstract}

\maketitle

\begin{figure}[b]
	\centering
	\includegraphics[width=8.6cm]{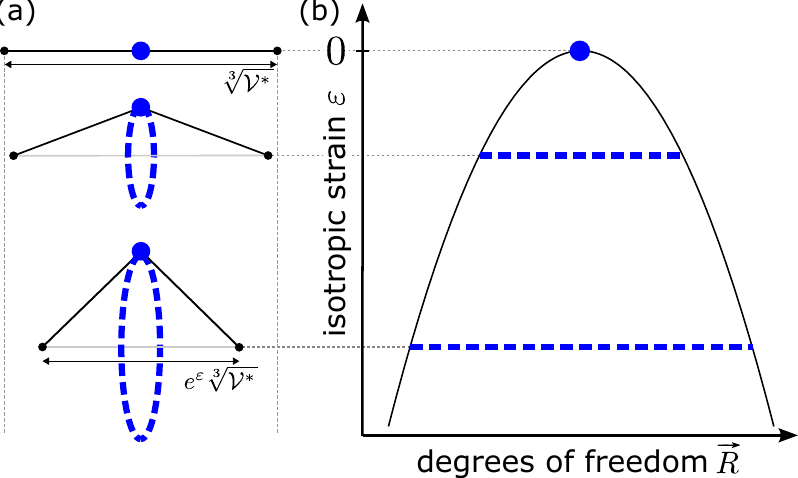}
	\caption{
		Variation of the accessible phase space volume $\Omega$ with isotropic strain $\varepsilon$ in the limit of infinitely stiff springs.	
		(a) Illustration of a simple two-spring system in 3D, of linear dimension $\sqrt[3]{\mathcal{V}}=e^\varepsilon\sqrt[3]{\mathcal{V}^\ast}$. The black nodes have fixed positions and the blue node is freely movable.
		(b) Starting at the athermal transition point at $\varepsilon=0$, upon isotropic compression ($\varepsilon<0$), $\Omega$ is expected to increase with the distance $\vert\varepsilon\vert$ to the athermal transition point. The abscissa represents the $N_\mathrm{dof}$-dimensional configuration space. 
		In both panels, blue dashed lines represent the accessible parts of the phase space.
	}
	\label{fig:overview-Omega}
\end{figure}

\begin{figure}[b]
	\centering
	\includegraphics[width=8.6cm]{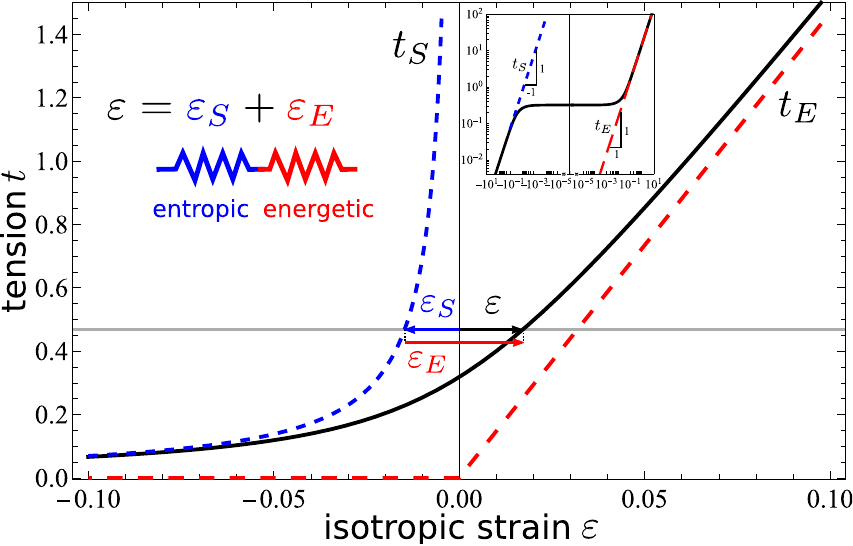}
	\caption{
		Analytical predictions for the isotropic tension $t$ depending on isotropic strain $\varepsilon$, shown for the athermal limit (red dashed line), the infinitely-stiff-spring limit (blue dashed line), and the general case of finite spring stiffness and temperature (black solid line).
		(inset) Same plot in log-log scaling.
	}
	\label{fig:overview-t-e}
\end{figure}

Understanding rigidity of amorphous materials still represents a fundamental challenge, which is relevant for questions ranging from the glass transition in inert systems \cite{Charbonneau2017} to solid-fluid transitions in living systems \cite{Lenne2022}.

In the athermal limit, i.e.\ at zero temperature, the rigidity of many systems including spring networks can be predicted using Maxwell-Calladine constraint counting \cite{Maxwell1864,Calladine1978,Liu2010,Lubensky2015},
where the number of constraints, $N_s$ (e.g.\ the number of springs), is compared to the number of degrees of freedom, $N_\mathrm{dof}$ (e.g.\ the node positions).
Roughly, if $N_s>N_\mathrm{dof}$, the system is expected to be rigid. Conversely, if $N_s<N_\mathrm{dof}$, the system is under-constrained and expected to be floppy.

However, it is known that even under-constrained systems can be rigidified \cite{Calladine1978,Alexander1998,Cui2019b,Merkel2019,Moshe2017,Damavandi2022c,Damavandi2022a,Zhang2022a}.
This occurs when the spring rest lengths are too short to accommodate an externally applied strain.
For example, in the simple spring network in Fig.~\ref{fig:overview-Omega}a, we illustrate what happens when progressively increasing the externally applied isotropic strain $\varepsilon$. 
Eventually, as $\varepsilon$ is increased above a critical value, here for convenience defined as the point of zero strain, $\varepsilon=0$, the springs necessarily need to be stretched.
At this point, a so-called state of self-stress (SSS) emerges \cite{Lubensky2015}. This means that a combination of virtual spring tensions does not result in any net forces on the (movable) nodes.
In previous work, we used the properties of this SSS to analytically derive the elastic properties in the athermal limit \cite{Merkel2019,Lee2022}.
Our approach applies not only to spring networks, but also to other systems such as polymer networks \cite{Onck2005,Broedersz2014,Licup2015,Sharma2016a} and vertex models for biological tissues \cite{Moshe2017,Merkel2018,Sussman2018,Merkel2019,Wang2020}.

While we analytically understand rigidity in the athermal limit, there are mostly only numerical results for the finite-temperature case, i.e.\ for \emph{thermal} under-constrained systems \cite{Plischke1998,Dennison2013,Wigbers2015,Sussman2018b,Woodhouse2018}. 
For instance, previous studies showed numerically that upon increasing temperature $T$ from the athermal rigidity transition point, the shear modulus $G$ scales as $G\sim T^{1/2}$ \cite{Dennison2013,Wigbers2015}. 
For certain spring networks, this scaling could be derived analytically using effective-medium theory (EMT) before \cite{Zhang2016a}.
There is also work treating a related abstract constraint satisfaction problem using the replica approach \cite{Urbani2023}, work studying these questions in isostatic or over-constrained systems \cite{Mao2015a,Zhang2016a,Rocklin2018a,Chen2023}, and work discussing the effect of singularities on under-constrained systems \cite{Mannattil2022}.
However, despite these advances, we still miss a general understanding for the elastic properties of thermal, under-constrained systems. In particular, previous EMT approaches are mean-field expansions, expected to be exact only for small network disorder. Also, previous approaches did not discuss the dependence of the elastic properties on isotropic and shear strain.

Here and in the companion paper \footnote{Reference to be inserted later.}, we develop a generic analytical theory for thermal, under-constrained systems in the vicinity of the athermal transition point.
In this paper, we first outline the derivation of the elastic system properties in the limit of infinitely stiff springs, where only entropic elasticity plays a role.
To then discuss the general case of finite spring stiffness and finite temperature (black solid line in Fig.~\ref{fig:overview-t-e}), we combine the stiff-spring results  (blue dashed line) with our earlier results for the athermal limit, where only energetic elasticity plays a role (red dashed line) \cite{Merkel2019,Lee2022}.
We provide an intuition for why the two limiting behaviors ``act in series'', i.e.\ entropic and energetic strains add up (illustration in Fig.~\ref{fig:overview-t-e}).
The rigorous derivations for all cases are in the companion paper \cite{Note1}.
Finally, we test our analytical results numerically, using simulations of randomly cut triangular networks. With only three parameters, whose values can be predicted from the microscopic network structure, our analytical results reproduce the observed behavior of tension $t$ and shear modulus $G$ over many orders of magnitude of varying isotropic strain $\varepsilon$, shear strain $\gamma$, and temperature $T$.

To discuss the key ideas of our approach, we focus here on an under-constrained network of $N_s$ linear springs that are connected at $N_\mathrm{node}$ nodes. The springs have lengths $L_i$ with equal spring constants $K$ and rest lengths $L_0$.
We substantially generalize this in the companion paper \cite{Note1}.
The system energy is:
\begin{equation}
	E = \frac{K}{2}\sum_{i=1}^{N_s}{(L_i-L_0)^2}.
\end{equation}
We use periodic boundary conditions in $D$ spatial dimensions with a periodic box of equal sides and with volume $\mathcal{V}$.  
Isotropic strain $\varepsilon$ is defined as linear strain, i.e.\ $\mathcal{V}=e^{D\varepsilon}\mathcal{V}^\ast$, where $\mathcal{V}^\ast$ is the system volume at the athermal transition point.
Shear strain $\gamma$ can be either pure or simple shear strain.

For example, for a squared periodic box in $D=2$ with simple shear strain $\gamma$, the length vector $\vecd{L}_i=(L_{i,x},L_{i,y})$ of a spring $i$ connecting the node with index $a_i$ to the node with index $b_i$ is given by:
\begin{equation}
	\begin{aligned}
		L_{i,x} &= R_{b_i,x} - R_{a_i,x} + (q_{i,x} + \gamma q_{i,y})e^\varepsilon \sqrt{\mathcal{V}^\ast} \\
		L_{i,y} &= R_{b_i,y} - R_{a_i,y} + q_{i,y}e^\varepsilon \sqrt{\mathcal{V}^\ast},
	\end{aligned}\label{eq:spring vector}
\end{equation}
where $\vecd{R}_a=(R_{a,x},R_{a,y})$ is the position of node $a$, and $\vecd{q}_i=(q_{i,x},q_{i,y})\in\mathbb{Z}^2$ is a periodicity vector, where $q_{i,x}$ ($q_{i,y}$) counts how often spring $i$ crosses the vertical (horizontal) periodic box boundary. The length of spring $i$ is then given by: $L_i=\vert \vecd{L}_i(\vecn{R},\varepsilon,\gamma)\vert$, where  $\vecn{R}$ denotes the $N_\mathrm{dof}=DN_\mathrm{node}$-dimensional vector that contains all node positions.
For convenience, we use dimensionless node positions $\vecd{r}_a:=\mathcal{V}^{-1/D}\vecd{R}_a=e^{-\varepsilon}(\mathcal{V}^\ast)^{-1/D}\vecd{R}_a$, i.e.\ the periodic box is rescaled to side lengths of one, for any value of $\varepsilon$.
The spring vectors are rescaled in a similar way, $\vecd{\ell}_i:=e^{-\varepsilon}\vecd{L}_i/L_0$, such that at the transition with $\varepsilon=0$ and $L_i=L_0$, we have dimensionless spring lengths of $\ell_i:=\vert\vecd{\ell}_i\vert=1$.
Thus, Eq.~\eqref{eq:spring vector} becomes:
\begin{equation}
	\begin{aligned}
		\ell_{i,x} &= (r_{b_i,x} - r_{a_i,x} + q_{i,x} + \gamma q_{i,y})\sqrt{\mathcal{V}^\ast}/L_0 \\
		\ell_{i,y} &= (r_{b_i,y} - r_{a_i,y} + q_{i,y})\sqrt{\mathcal{V}^\ast}/L_0.
	\end{aligned}
\end{equation}
Hence, the dimensionless spring lengths $\ell_i(\vecn{r},\gamma)$ do not depend on isotropic strain $\varepsilon$ when expressed in terms of the dimensionless node positions $\vecn{r}$.

In the athermal limit, the elastic properties arise from purely energetic interactions, and we have derived their analytical expressions before \cite{Merkel2019,Lee2022}. Briefly, for small isotropic and shear strain, the system is rigid for $\varepsilon+b_\varepsilon\gamma^2>0$, where $b_\varepsilon$ is a parameter that depends on the microscopic network structure. In the rigid regime close to the transition, isotropic tension $t_E$ (red dashed line in Fig.~\ref{fig:overview-t-e}) and shear modulus $G_E$ are given by \cite{Note1}:
\begin{align}
	t_E &= \kappa_E\big(\varepsilon+b_\varepsilon\gamma^2\big) \label{eq:tE}\\
	G_E &= 2Db_\varepsilon\kappa_E\big(\varepsilon+3b_\varepsilon\gamma^2\big),
\end{align}
where $\kappa_E$ is another parameter that depends on the microscopic network structure.

In the thermal case, the configurational degeneracy of the system becomes important (Fig.~\ref{fig:overview-Omega}).
To discuss its influence, we first consider the stiff-spring limit, $K\rightarrow\infty$, which implies a hard constraint for all spring lengths: $L_i=L_0$, which translates into dimensionless units as $\ell_i=e^{-\varepsilon}$, which becomes $\ell_i=1-\varepsilon$ to linear order in $\varepsilon$.
To obtain an expression for the accessible phase space volume $\Omega$, we Taylor-expand all dimensionless spring lengths around the transition.
Assuming that, up to global translation, the node coordinates at the transition point, $\vecn{r}\,^\ast$, are uniquely defined, we denote the deviation from these positions by $\Delta\vecn{r}:=\vecn{r}-\vecn{r}\,^\ast$. For small $\Delta\vecn{r}$, we expand $\Delta\ell_i(\vecn{r},\gamma=0)$ to second order:
\begin{equation}
	-\varepsilon\equiv \Delta\ell_i = \sum_{n=1}^{N_\mathrm{dof}} C_{in}\Delta r_n  + \frac{1}{2}\sum_{m,n=1}^{N_\mathrm{dof}}M_{imn}\Delta r_m\Delta r_n,\label{eq:Taylor-expansion}
\end{equation}
Here, we introduced $\Delta\ell_i:=\ell_i-1$, which becomes $\Delta\ell_i=-\varepsilon$ in the stiff-spring limit where $L_i=L_0$.
The $\Delta r_n$ with $n=1,\dots,N_\mathrm{dof}$ are the components of $\Delta\vecn{r}$.
We introduced the compatibility matrix $C_{in}:=\partial\ell_i/\partial r_n$ and $M_{imn}:=\partial^2\ell_i/\partial r_m\partial r_n$, both evaluated at the transition point.
The derivation for nonzero $\gamma$ is in the companion paper \cite{Note1}.

The accessible phase space corresponds to the set of positions $\vecn{r}$ where Eq.~\eqref{eq:Taylor-expansion} holds. To derive an expression for its volume $\Omega$, we express Eq.~\eqref{eq:Taylor-expansion} in terms of the eigenmodes of $C_{in}$, by performing a singular-value decomposition on $C$: $C_{in}=\sum_{p=1}^{N_s}{U_{ip}s_pV_{np}}$, with the singular values $s_p$ and orthogonal square matrices $U$ and $V$. 
Multiplying Eq.~\eqref{eq:Taylor-expansion} by $U_{ip}$ and summing over $i$, we obtain for $1\leq p\leq N_s$:
\begin{align}
	-\varepsilon\tilde{w}_p = s_p\Delta \tilde{r}_p + &\frac{1}{2}\sum_{q,r=N_s}^{N_\mathrm{dof}}\tilde{M}_{pqr}\Delta\tilde{r}_q\Delta\tilde{r}_r \label{eq:Taylor-expansion-em-2}
\end{align}
where we set $\tilde{w}_p:=\sum_{i=1}^{N_s}{U_{ip}}$, $\Delta\tilde{r}_p:=\sum_{n=1}^{N_\mathrm{dof}} V_{np}\Delta r_n$, and $\tilde{M}_{pqr} := \sum_{i=1}^{N_s}\sum_{m,n=1}^{N_\mathrm{dof}} U_{ip}M_{imn}V_{mq}V_{nr}$.
Each SSS corresponds to a singular mode with $s_p=0$ \cite{Lubensky2015}. Here we focus on the generic case, where a single SSS is created at the transition point \cite{Lubensky2015,Merkel2019}. Sorting the $s_p$ in decreasing order, we thus have that $s_p>0$ for $p<N_s$ and $s_{N_s}=0$. 
In the companion paper, we show that the terms $\sim\Delta\tilde{r}_q\Delta\tilde{r}_r$ with $q<N_s$ or $r<N_s$ are of higher than linear order in $\varepsilon$ and can thus be neglected \cite{Note1}.

Eq.~\eqref{eq:Taylor-expansion-em-2} allows us to discuss the shape of the accessible phase space volume. The first $N_s-1$ equations in Eq.~\eqref{eq:Taylor-expansion-em-2} only fix the values of the $\Delta\tilde{r}_p$ with $p<N_s$. Meanwhile, the last equation, $p=N_s$, puts a constraint on the $\Delta\tilde{r}_q$ with $q\geq N_s$, where $s_{N_s}=0$ and $\tilde{M}_{N_sqr}$ is a positive semi-definite square matrix for $q,r=N_s,\dots,N_\mathrm{dof}$ (the first out of the three indices of $\tilde{M}_{N_sqr}$ being fixed) \cite{Note1}. Thus, Eq.~\eqref{eq:Taylor-expansion-em-2} with $p=N_s$ describes the surface of a hyper-ellipsoid, whose half axes scale as $\sim\sqrt{-\varepsilon}$.
Some of the eigenvalues of $\tilde{M}_{N_sqr}$ are zero; for instance those that correspond to a global translation of all nodes. We denote the number of non-zero eigenvalues of $\tilde{M}_{N_sqr}$ by $N_\mathrm{1st}^\ast$, thus the ellipsoid is $N_\mathrm{1st}^\ast$-dimensional. 
These $N_\mathrm{1st}^\ast$ eigenvalues correspond to what we call \emph{first-order zero modes} at the transition \cite{Note1}; collective motions of the nodes that change the spring lengths \emph{not} to first order \emph{but} to second order.
In the example shown in Fig.~\ref{fig:overview-Omega}a, we have $N_\mathrm{1st}^\ast=2$, corresponding to the two infinitesimal zero modes at the transition, and the accessible phase-space volume is a circle (blue dashed curves), i.e.\ the surface of a 2-dimensional hyper-ellipsoid.
In the general case, the surface scales as $\sim (-\varepsilon)^{(N_\mathrm{1st}^\ast-1)/2}$, and together with an additional prefactor of $\sim 1/\sqrt{-\varepsilon}$ due to the constraint on the spring lengths \cite{Note1}, we obtain for the accessible phase space volume:
\begin{equation}
	\Omega \sim (-\varepsilon)^{\frac{N_\mathrm{1st}^\ast-2}{2}}.
\end{equation}
The free energy consists only of an entropic contribution $F_S=-k_BT\log{\Omega}=-k_BT([N_\mathrm{1st}^\ast-2]/2)\log{[-(\varepsilon+b_\varepsilon\gamma^2)]}$, where $k_B$ denotes the Boltzmann constant, and we neglected constant offsets. Also, we have included the dependency on $\gamma$, obtained in the companion paper \cite{Note1}.
Note that in the stiff-spring limit, the network cannot be arbitrarily strained -- only strains such that $\varepsilon+b_\varepsilon\gamma^2<0$ are allowed.
We obtain for purely entropic tension $t_S=(\partial F_S/\partial\varepsilon)/D\mathcal{V}$ and shear modulus $G_S=(\partial^2 F_S/\partial\gamma^2)/\mathcal{V}$:
\begin{align}
	t_S &= -\frac{\kappa_ST}{\varepsilon+b_\varepsilon\gamma^2} \label{eq:tS}\\
	G_S &= -2Db_\varepsilon\kappa_ST\frac{\varepsilon-b_\varepsilon\gamma^2}{(\varepsilon+b_\varepsilon\gamma^2)^2},
\end{align}
with $\kappa_S=k_B(N_\mathrm{1st}^\ast-2)/2D\mathcal{V}^\ast$.
Hence, not only energetic rigidity \cite{Merkel2019,Lee2022}, but also entropic rigidity is determined by the properties of the SSS that is created at the athermal transition point.

For $\gamma=0$, the scaling of the tension $t_S$ is plotted in Fig.~\ref{fig:overview-t-e} as a function of $\varepsilon$ (blue dashed line).
This tension scaling is consistent with known results for some under-constrained systems, including freely jointed chains \cite{Rubinstein2003} or incompressible membranes without bending rigidity \cite{Durand2022} close to their respective fully stretched states.

\begin{figure*}
	\centering	\includegraphics[width=18cm]{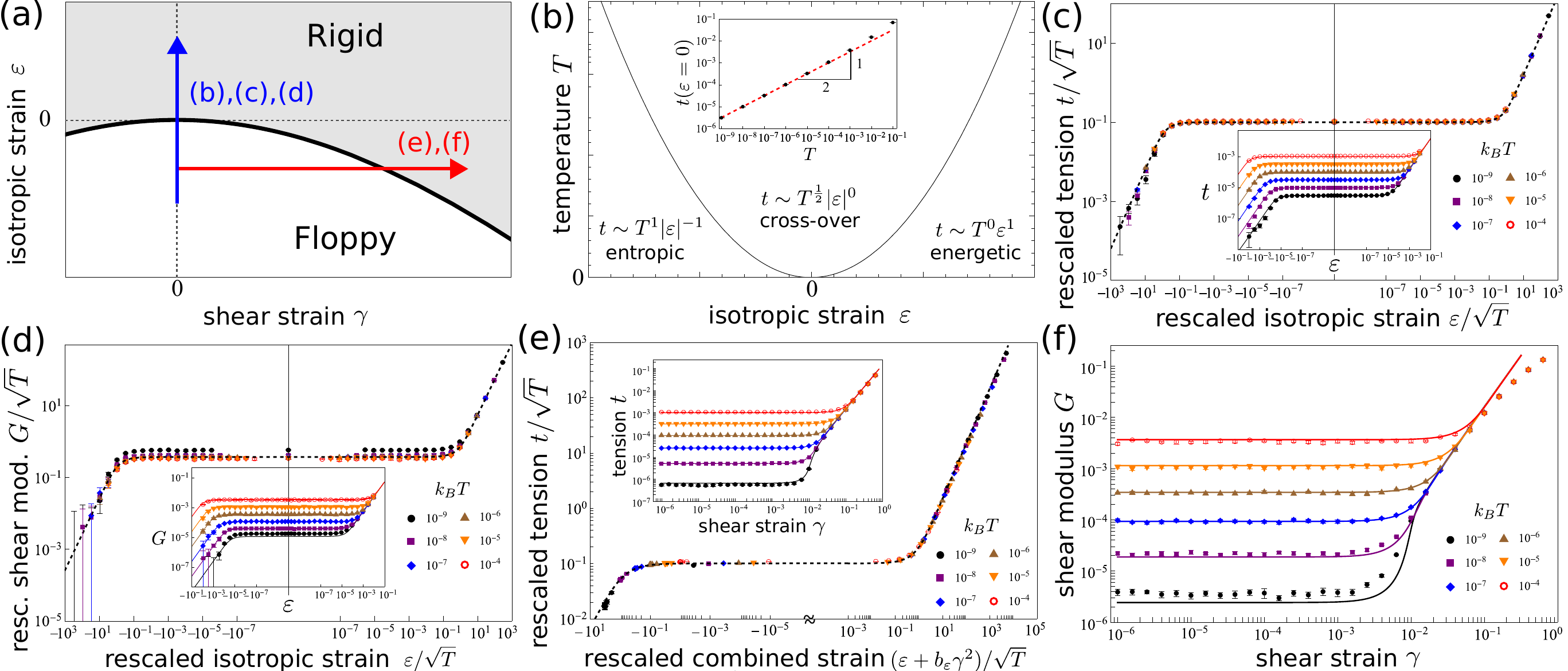}
	\caption{ Numerical results. 
		(a) Schematic phase diagram: athermal rigid (gray) and floppy (white) regimes, depending on isotropic and shear strain. We used two different strain protocols:  Varying isotropic strain at zero shear strain, $\gamma=0$ (blue arrow; panels b, c, d), and varying shear strain, $\gamma>0$, with fixed $\varepsilon=-9.15\times10^{-5}$ (red arrow; panels e, f).
		(b) For varying isotropic strain we find three scaling regimes. (b inset)  For $\varepsilon=0$, we numerically confirm the scaling $t\sim T^{1/2}$.  The red dashed line is a fit to the curve $t=\sqrt{\kappa_E\kappa_ST}$ for $T \leq 10^{-6}$.
		(c, d) When plotting isotropic tension $t$ (panel c) or shear modulus $G$ (panel d) versus isotropic strain $\varepsilon$, a rescaling by $\sqrt{T}$ leads to a collapse of the curves for the different temperatures. (c,d inset) Same data without rescaling.
		(e) When plotting isotropic tension $t$ versus the combined strain, $\varepsilon+b_\varepsilon\gamma^2$, a rescaling by $\sqrt{T}$ leads to a collapse of the curves for the different temperatures.
		(e inset) Same data without rescaling.
		(f) Shear modulus $G$ versus shear strain $\gamma$.
 		All curves in panels c-f and their insets are predictions from Eqs.~\eqref{eq:t-e-g} and \eqref{eq:G-e-g} using a single parameter set $(\kappa_E,\kappa_S,b_\varepsilon)$ (Appendix~B), whose values can be analytically computed from the network microstructure (Appendix~C). 
 		}
	\label{fig:numerical-results}
\end{figure*}

We now discuss the general case of \emph{finite} spring stiffness $K$ and \emph{finite} temperature $T$.
To get an intuition for the interplay between entropic and energetic elasticity, we consider boundary conditions that ensure $\gamma=0$ and constant tension $t>0$ (horizontal gray line in Fig.~\ref{fig:overview-t-e}).
In the thermal stiff-spring limit, $K\rightarrow\infty$, this leads to a negative strain $\varepsilon_S<0$ (blue arrow), which is according to Eq.~\eqref{eq:tS}: $\varepsilon_S = -\kappa_ST/t$ (blue dashed line).
When we now allow for a finite spring constant $K$, the springs additionally stretch, increasing the total strain by some amount $\varepsilon_E$ to the total strain $\varepsilon=\varepsilon_S+\varepsilon_E$ (red and black arrows, respectively).
For sufficiently small temperatures, the network geometry is on average not very different from the athermal network. So, using Eq.~\eqref{eq:tE}, we have $\varepsilon_E = t/\kappa_E$ (red dashed line).
Thus, the total strain depends on tension as  (black solid line)
\begin{equation}
	\varepsilon = -\frac{\kappa_ST}{t} + \frac{t}{\kappa_E}.\label{eq:e-t}
\end{equation}
In other words, entropic and an energetic rigidity act in series (Fig.~\ref{fig:overview-t-e}).
The rigorous analysis in our companion paper confirms that this intuitive explanation leads to the exact result for $N_\mathrm{1st}^\ast\gg1$ \cite{Note1}, a limit that we expect many physically relevant systems to be in.

Inverting Eq.~\eqref{eq:e-t} we find:
\begin{equation}
	t = \frac{\kappa_E}{2}\Big(\varepsilon + b_\varepsilon\gamma^2 + \big\vert\varepsilon + b_\varepsilon\gamma^2\big\vert\sqrt{1+\theta}\Big), \label{eq:t-e-g}
\end{equation}
where $\theta:=4\kappa_ST/[\kappa_E(\varepsilon + b_\varepsilon\gamma^2)^2]$.
Again, the $\gamma$ dependency included here is obtained in the companion paper \cite{Note1}.
We find for the shear modulus \cite{Note1}:
\begin{equation}
	G = 2Db_\varepsilon t\left[ 1 + \frac{2b_\varepsilon\gamma^2}{\vert\varepsilon + b_\varepsilon\gamma^2\vert\sqrt{1+\theta}}\right].\label{eq:G-e-g}
\end{equation}
Our approach also provides a simple explanation for the previously observed $G\sim T^{1/2}$ scaling for $\varepsilon=0$ and $\gamma=0$. In this limit, we find from Eqs.~\eqref{eq:t-e-g} and \eqref{eq:G-e-g} that $t=\sqrt{\kappa_E\kappa_ST}$ and $G=2Db_\varepsilon t$.
The relation for $t$ can also be obtained immediately from Eq.~\eqref{eq:e-t} with $\varepsilon=0$.

More generally, we find three scaling regimes, which we discuss here for $\gamma=0$ (Fig.~\ref{fig:numerical-results}b and inset to Fig.~\ref{fig:overview-t-e}).
(i) For $\theta\ll1$ and $\varepsilon<0$, the system behaves like in the stiff-spring limit with $G\sim t\sim T^1\vert\varepsilon\vert^{-1}$.
(ii) For $\theta\ll1$ and $\varepsilon>0$, the system behaves like in the athermal regime with $G\sim t\sim T^0\varepsilon^1$.
(iii) For $\theta\gg1$, the system shows the scaling $G\sim t\sim T^{1/2}\varepsilon^0$.
This is remarkably analogous to previous results \cite{Zhang2016a} when replacing strain by the network connectivity. We discuss possible reasons for this in the companion paper \cite{Note1}.

We tested our analytical results using Monte-Carlo (MC) simulations of a 2D under-constrained randomly cut triangular network in periodic boundary conditions (Appendix~A).
We first set shear strain to zero, $\gamma=0$, varying only isotropic strain $\varepsilon$ and temperature $T$ (blue arrow in Fig.~\ref{fig:numerical-results}a, Fig.~\ref{fig:numerical-results}b-d).
We confirmed the $t\sim T^{1/2}$ scaling for $\varepsilon=0$ (inset to Fig.~\ref{fig:numerical-results}b). 
Furthermore, Eqs.~\eqref{eq:t-e-g} and \eqref{eq:G-e-g} predict that for $\gamma=0$ the tension-strain and shear-modulus-strain curves respectively collapse when rescaling each of $t$, $G$, and $\varepsilon$ by $\sqrt{T}$, which we confirm numerically in Fig.~\ref{fig:numerical-results}c,d. The theoretical predictions are shown as dashed and solid curves, where the values of $\kappa_E$, $\kappa_S$, and $b_\varepsilon$ were obtained from fits to a subset of the data
(Appendix~B).
We verified that all three values match those analytically predicted directly from the network structure (Appendix~C).

We further tested our analytical results for varying shear strain, $\gamma>0$, fixing isotropic strain $\varepsilon$ (red arrow in Fig.~\ref{fig:numerical-results}a, Fig.~\ref{fig:numerical-results}e,f).
For the tension $t$, Eq.~\eqref{eq:t-e-g} again implies a scaling collapse onto a theoretical curve, which we confirm numerically in Fig.~\ref{fig:numerical-results}e. The solid and dashed lines in Fig.~\ref{fig:numerical-results}e and inset indicate again the prediction according to Eq.~\eqref{eq:t-e-g}, using the same parameter values as before (Appendix B).
Similarly, the shear modulus data follows the analytical prediction in Eq.~\eqref{eq:G-e-g} (Fig.~\ref{fig:numerical-results}f). Deviations in the shear modulus at the lowest temperature are consistent with slightly insufficient equilibration, and deviations at larger shear strain are likely due to higher-order terms.

In conclusion, we have developed a generic analyical theory for the elastic properties of thermal, under-constrained systems, and their behavior under isotropic and shear strain. Our results hold independent of network disorder. 
To obtain our analytical results, a few assumptions were required. Specifically, our analytical results hold only in the limit of small strain $\varepsilon,\gamma$ and temperature $T$. Moreover, for simplicity, we assumed that there are no SSS in the floppy regime (implying no residual stresses in this regime), and only a single SSS forms at the transition. All technical assumptions made are discussed in the companion paper \cite{Note1}. In future work, it will be interesting to study how much each of these assumptions can be alleviated.
We expect this work to unify the physics of a broad class of materials, including polymer fibers, polymer networks, membranes, and vertex models for biological tissues.

\begin{acknowledgments}
We thank Chris Santangelo, Jen Schwarz, and Manu Mannattil for fruitful discussions.
We thank the Centre Interdisciplinaire de Nanoscience de Marseille (CINaM) for providing office space.
The project leading to this publication has received funding from France 2030, the French Government program managed by the French National Research Agency (ANR-16-CONV-0001), and from the Excellence Initiative of Aix-Marseille University - A*MIDEX.
%
\end{acknowledgments}

\bibliography{references.bib}

\appendix
\section{End Matter}
\subsection{A - Simulations}
The network was initialized with $40\times 40$ nodes with connectivity $z=3.2$ \cite{Broedersz2011b,Shivers2019,Arzash2019,Lee2022}, were we used spring constant $K=1$ and rest length $L_0=1$. 
Dangling springs and isolated islands were removed and were not counted towards $z$, and the network was shear-stabilized \cite{Merkel2019,Lee2022}.
We used simple shear strain $\gamma$, and in a slight deviation from the analytical part, in the simulations we defined the isotropic strain such that $\mathcal{V}=(1+\varepsilon)^D\mathcal{V}^\ast$. The difference is only of higher-order in $\varepsilon$, and thus not relevant here.
We used the Metropolis–Hastings algorithm \cite{Frenkel2002} with $10^{10}$ MC steps, where in each step we moved a single, randomly selected node. The node displacement was chosen such that the transition probability was $\approx0.5$. We computed the network stress tensor as the time-averaged force dipole density of the individual springs \cite{Batchelor1970} and subtracted the ideal gas contribution \cite{deMiguel2006,Dennison2013,Tuckerman2010}.
The shear modulus was computed from the shear stress using a central finite difference quotient with $\Delta\gamma = 2.5\times 10^{-3}$.
Results were averaged over 10 different simulation runs for each parameter set $(\varepsilon,\gamma,T)$.

\subsection{B - Parameter fits}
We extracted $\kappa_E\approx 0.161$ from the tension data for the athermal limit, $t(\varepsilon, \gamma=0, T=0)$, using a linear fit (compare Eq.~\eqref{eq:tE}). The value of $\kappa_S/k_B\approx 0.0629$ was then computed from $\kappa_E\kappa_S/k_B\approx 0.0101$, obtained from the fit in the inset to Fig.~\ref{fig:numerical-results}b. The parameter $b_\varepsilon\approx0.835$ describing the interaction between isotropic and shear strain was obtained from the averaged ratio $G/t$ at $(\varepsilon,\gamma)=(0,0)$ for $ 10^{-8}\leq T \leq 10^{-5}$.
These three parameters was then used to predict the behavior of tension $t$ and shear modulus $G$ for all studied tuples $(\varepsilon,\gamma,T)$ (Fig.~\ref{fig:numerical-results}c-f).

\subsection{C - Direct parameter prediction from the network structure}
While we obtained the values used for $\kappa_E$, $\kappa_S$, and $b_\varepsilon$ from fits to numerical data, our theory predicts almost the same values directly from the network structure.
To obtain a prediction for $\kappa_E$, we numerically identified $\tilde{w}_{N_s} \approx 22.785$ from the $U$ matrix obtained by a singular-value decomposition of the compatibility matrix $C_{in}$. With the critical network area $\mathcal{V}^\ast = 1658$, this gives $\kappa_E = \tilde{w}_{N_s}^2/D\mathcal{V}^\ast\approx 0.156$.

To obtain a prediction for $\kappa_S$, we determined
$N_\mathrm{1st}^\ast = 433$ as the number of positive eigenvalues of the matrix $\tilde{M}_{N_sqr}$ at the athermal transition. This gives $\kappa_S/k_B=(N_\mathrm{1st}^\ast-2)/2D\mathcal{V}^\ast \approx 0.0650$. 
While the deviation of measured $\kappa_E$ is $\approx 3\%$, the deviation of the measured product $\kappa_E\kappa_S$ is only $\approx 0.1\%$.

Finally, we also predicted the parameter $b_\varepsilon$ using the formula in Appendix~C of the companion paper~\cite{Note1}, and we found $b_\varepsilon\approx0.850$, which corresponds to a deviation of less than $2\%$ from the fitted value.
Taken together, we find only small deviations in the percentage range, which may come from higher-order terms in $\varepsilon$ and $\gamma$.

\end{document}